\documentclass[floatfix,showpacs,amsmath,amsfonts,amssymb,aps,twocolumn,superscriptaddress,prb,10pt]{revtex4-1}
\usepackage[pdftex]{graphicx}
\usepackage[eulergreek]{sansmath}
\usepackage[caption=false,position=top]{subfig}
\usepackage{array}
\usepackage{bm}
\usepackage[colorlinks=true,citecolor=blue,linkcolor=blue,urlcolor=blue]{hyperref}
\usepackage{dcolumn}
\usepackage{mathtools}

\DeclarePairedDelimiter\bra{\langle}{\rvert}
\DeclarePairedDelimiter\ket{\lvert}{\rangle}
\DeclarePairedDelimiterX\braket[2]{\langle}{\rangle}{#1 \delimsize\vert #2}

\DeclareMathAlphabet{\mathbfsf}{\encodingdefault}{\sfdefault}{bx}{n}

\newcolumntype{.}{D{.}{.}{-1}}

\newcommand{\vect}[1]{\mathbf{#1}}

\newcommand{\kf}[0]{k_\mathrm{F}}

\newcommand{\figref}[1]{Fig.~\ref{#1}}

\newcommand{\neweqnline}{\nonumber\\}
\newcommand{\secref}[1]{Section~\ref{#1}}
\newcommand{\eqnref}[1]{Eq.~(\ref{#1})}

\newcommand{\vecgrk}[1]{\boldsymbol{#1}}
\newcommand{\kfup}[0]{k_{\uparrow \mathrm{F}}}
\newcommand{\kfdn}[0]{k_{\downarrow \mathrm{F}}}
\newcommand{\kfsigma}[0]{k_{\sigma \mathrm{F}}}

\newcommand{\Nup}[0]{N_\uparrow}
\newcommand{\Ndn}[0]{N_\downarrow}
\newcommand{\Nc}[0]{N_\mathrm{c}}
\newcommand{\Nsigma}[0]{N_\sigma}
\newcommand{\Sup}[0]{L_\uparrow}
\newcommand{\Sdn}[0]{L_\downarrow}
\newcommand{\Ssigma}[0]{L_\sigma}
\newcommand{\Efup}[0]{E_{\uparrow \mathrm{F}}}
\newcommand{\Efdn}[0]{E_{\downarrow \mathrm{F}}}
\newcommand{\Efsigma}[0]{E_{\sigma \mathrm{F}}}

\newcommand{\Eb}[0]{E_{\mathrm{b}}}
\newcommand{\nuup}[0]{\nu_{\uparrow}}
\newcommand{\nudn}[0]{\nu_{\downarrow}}
\newcommand{\nuc}[0]{\nu_{\mathrm{c}}}
\newcommand{\nusigma}[0]{\nu_{\sigma}}
\newcommand{\state}[2]{\ket*{#1 ; #2}}
\newcommand{\brastate}[2]{\bra*{#1 ; #2}}
\newcommand{\omegaD}[0]{\omega_\mathrm{D}}
\newcommand{\kD}[0]{k_\mathrm{D}}
\newcommand{\Kup}[0]{\mathbfsf{K}_\uparrow}
\newcommand{\Kdn}[0]{\mathbfsf{K}_\downarrow}
\newcommand{\Ksigma}[0]{\mathbfsf{K}_\sigma}

\begin{document}
\title{Multi-particle instability in a spin-imbalanced Fermi gas}

\author{T.M.~Whitehead}
\affiliation{Cavendish Laboratory, J.J. Thomson Avenue, Cambridge, CB3 0HE, 
United Kingdom}
\author{G.J.~Conduit}
\affiliation{Cavendish Laboratory, J.J. Thomson Avenue, Cambridge, CB3 0HE, 
United Kingdom}
\date{\today}

\begin{abstract}
Weak attractive interactions in a spin-imbalanced Fermi gas induce a 
multi-particle instability, binding multiple fermions together.  The maximum 
binding energy per particle is achieved when the ratio of the number of up- and 
down-spin particles in the instability is equal to the ratio of the up- and 
down-spin densities of states in momentum at the Fermi surfaces, to utilize the 
variational freedom of all available momentum states.  We derive this result 
using an analytical approach, and verify it using exact diagonalization.  The 
multi-particle instability extends the Cooper pairing instability of balanced 
Fermi gases to the imbalanced case, and could form the basis of a many-body 
state, analogously to the construction of the Bardeen--Cooper--Schrieffer 
theory of superconductivity out of Cooper pairs.
\end{abstract}

\maketitle

\section{Introduction}

Attractive interactions have a long and noteworthy history as the progenitors 
of strongly correlated states.  One of the earliest yet most profound insights 
was that attractive interactions between up- and down-spin electrons may induce 
a pairing instability, resulting in the formation of Cooper 
pairs~\cite{Cooper56}.  These Cooper pairs then form the basis of the many-body 
Bardeen--Cooper--Schrieffer (BCS) theory of 
superconductivity~\cite{Bardeen57,Bardeen57a}. Furthermore, even when there are 
unequal numbers of up- and down-spin particles in a system, 
Fulde~and~Ferrell~\cite{Fulde64} and, separately, 
Larkin~and~Ovchinnikov~\cite{Larkin65}~(FFLO) showed that it is still 
energetically favorable for up- and down-spin particles from their respective 
Fermi surfaces to form Cooper pairs, leading to a strongly correlated 
superconducting phase in spin-imbalanced Fermi 
gases~\cite{Casalbuoni04,Bowers02}. However, the density of states in momentum 
at the Fermi surface of the majority-spin particles is greater than that of the 
minority-spin species, so the number of bound pairs that can exist is limited 
by the number of minority-spin particles, leaving many of the majority-spin 
particles at their Fermi surface
unpaired and so uncorrelated. We propose a multi-particle instability that 
involves multiple majority-spin particles for each minority-spin particle, 
allowing us to utilize all of the potential of the majority-spin particles for 
contributing correlation energy. We find that the number of particles involved 
in the instability per species is proportional to the density of states in 
momentum at their respective Fermi surfaces. The multi-particle instability has 
more binding energy per particle than a Cooper pair, so could replace the 
Cooper pair as the building block of a superconducting state in spin-imbalanced 
Fermi gases.

The prototypical experimental realization of an imbalanced Fermi gas is
electrons in an external magnetic field. Most superconductors are destroyed
by an external magnetic field, reverting to the normal phase. However some 
materials, including 
CeCoIn$_{5}$~\cite{Bianchi03} and
\mbox{$\kappa$-(BEDT-TTF)$_{2}$Cu(NCS)$_{2}$~\cite{Mayaffre14}}, which are 
superconducting
at zero magnetic field, with increasing field undergo a phase transition
into an exotic second superconducting state, before a further transition into 
the
normal phase.  Other materials, including ErRh$_4$B$_4$~\cite{Prozorov08} and 
ErNi$_2$B$_2$C~\cite{Canfield96}, display overlap of ferromagnetism and 
superconductivity at zero applied field, and it has recently been suggested 
that Bi$_2$Sr$_2$CaCu$_2$O$_{8+x}$ exhibits some characteristics of an 
FFLO-like phase in the pseudogap regime~\cite{Hamidian16}. Further possible 
realizations of FFLO superconductivity in spin-imbalanced Fermi gases include 
an ultracold atomic gas of
fermions trapped in one dimension that displays a transition between
superconducting phases~\cite{Liao10}, or a spin-orbit coupled superconductor
with imbalanced Fermi surfaces~\cite{Zhang13,Lo14}. However, the exotic
superconducting state has not been fully characterized in any of these
systems, leaving the true nature of the ground state an open question.

We follow the prescription of Cooper~\cite{Cooper56} to study a multi-particle
instability on top of the Fermi surfaces. Working in second quantization 
notation, we construct a trial 
wavefunction for a multi-particle instability of several majority-spin particles
binding to a (potentially smaller) number of minority-spin particles to make 
the binding energy per particle larger than
for a Cooper pair. The optimal ratio for the number of majority- to 
minority-spin particles
is found to be the ratio of the densities of states in momentum at their 
respective Fermi surfaces.

To verify our multi-particle instability we analyze the system with exact
diagonalization. We confirm that our second quantized wave function captures
the crucial correlations of the exact solution, expose additional insights into 
the structure of the wavefunction, and verify our conclusion
that the optimal number of particles in the instability is set by the ratio
of the densities of states in momentum.

\section{Theory}
\label{sec:Theory}

\begin{figure*}[t]
\centering
\subfloat[$(\Nup,\Ndn)=(1,1)$ Cooper pair instability]{
\includegraphics[width=0.4\linewidth]{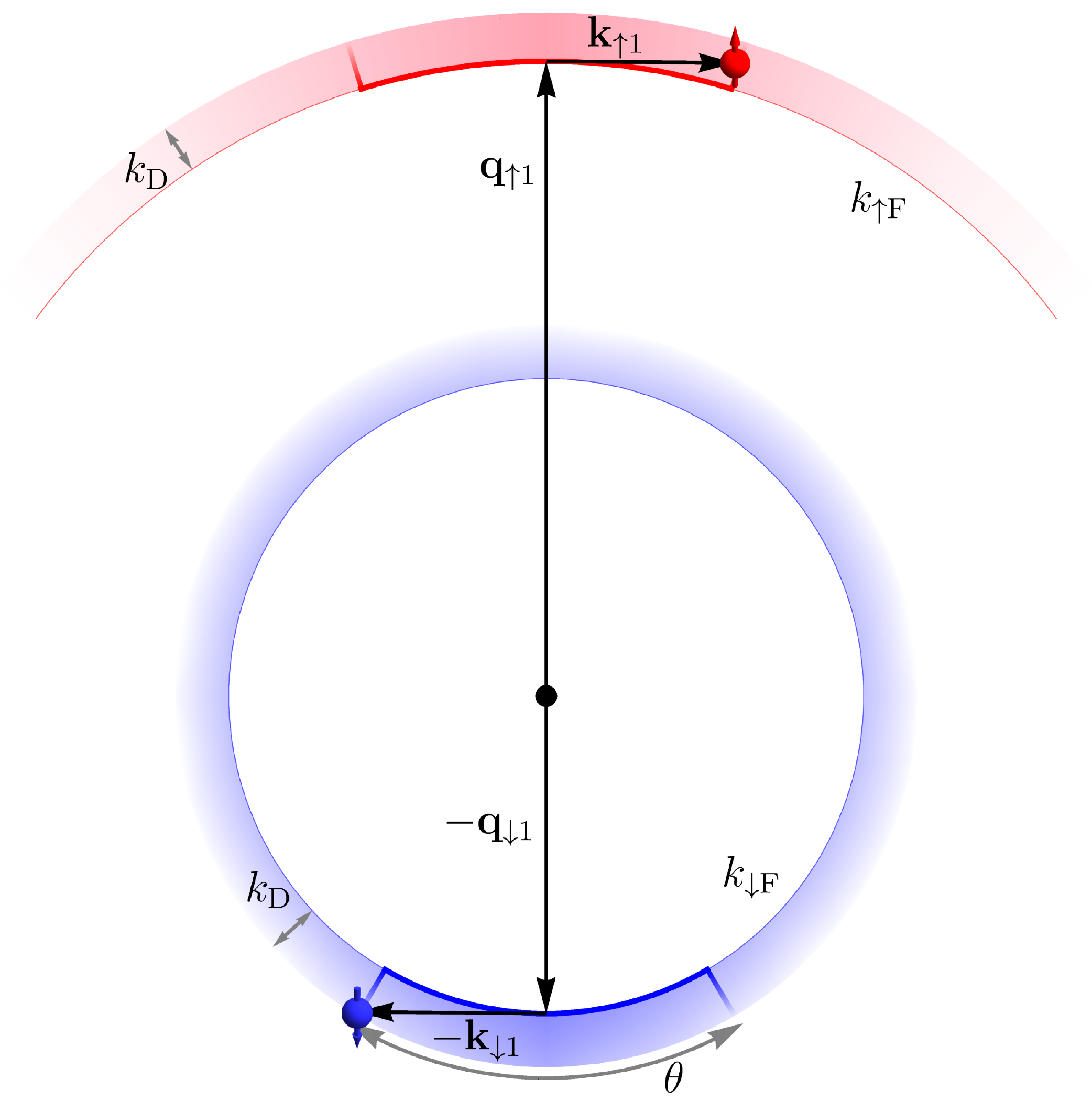}
\label{fig:IdealCirclesFFLO}
}\hspace{1.5cm}
\subfloat[$(\Nup,\Ndn)=(2,1)$ multi-particle instability]{
\includegraphics[width=0.4\linewidth]{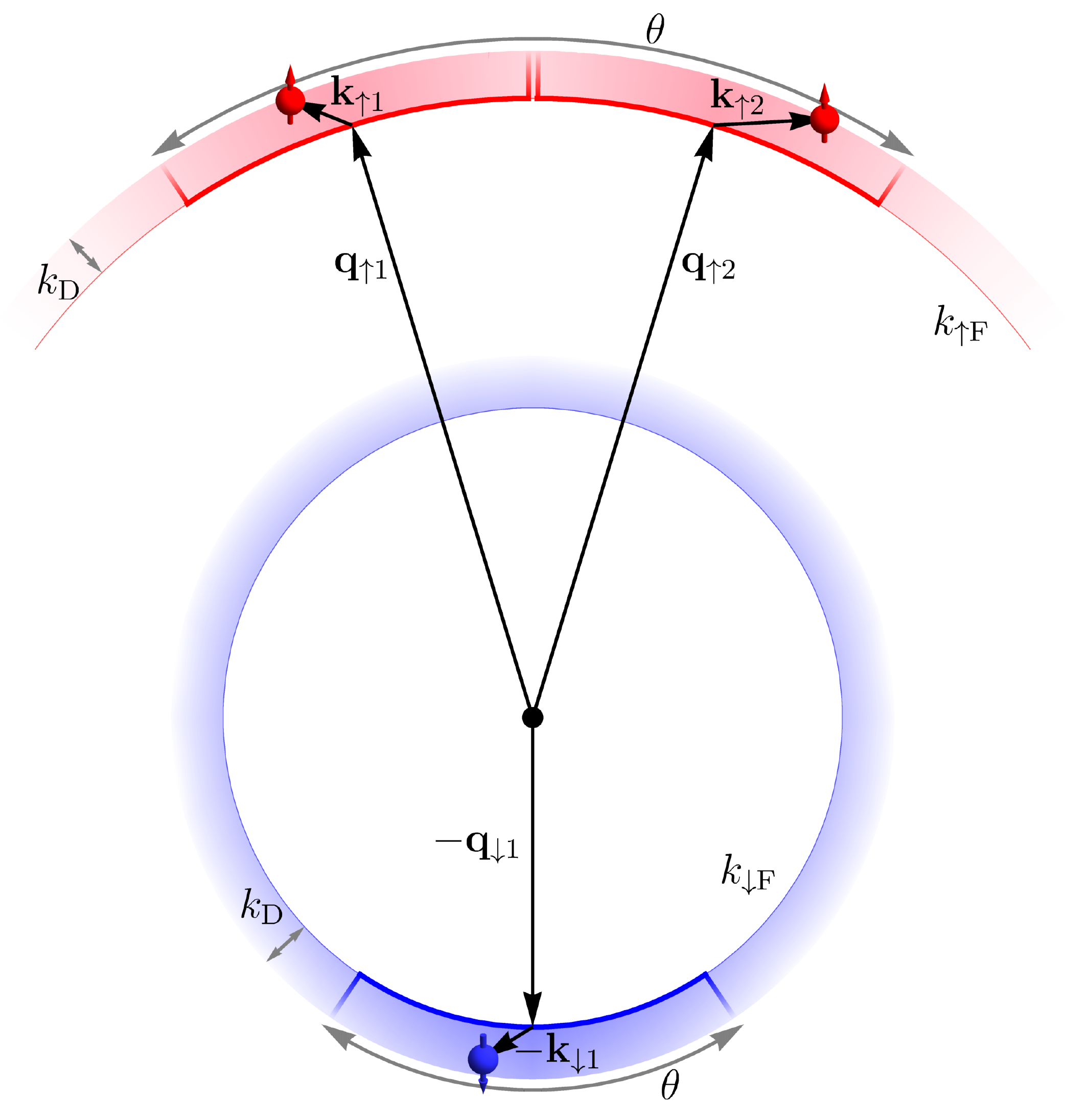}
\label{fig:IdealCircles}
}
\caption{(Color online) Idealized representation of the spin-imbalanced system 
showing Fermi surfaces for the down- (light blue circle) and up-spin (light red 
fragment of circle) species, with shaded areas denoting the allowed momentum 
states extending outwards by the Debye momentum $\kD$. Also shown are Fermi 
surface arcs, bounded by thick blue lines for the down-spin species and thick 
red lines for the up-spin species, for (a) the simplest instability of one 
up-spin and one down-spin particle, and (b) a proposed multi-particle 
instability with $(\Nup,\Ndn)=(2,1)$, indicating the momentum states used in 
the trial wavefunctions.}
\label{fig:Circles}
\end{figure*}

To explore the possibility of the multi-particle instability we study a 
two-spin fermionic system with an attractive contact interaction at zero 
temperature.  The BCS Hamiltonian takes the form
\begin{align}
\hat{H}=\sum_{\sigma,\vect{k}} \xi_{\sigma \vect{k}}  
c^\dagger_{\sigma\vect{k}} c_{\sigma\vect{k}} 
- g \!\sum_{\vect{k},\vect{k}',\vect{q}}\! 
c^\dagger_{\uparrow(\vect{q}-\vect{k})}c^\dagger_{\downarrow\vect{k}}
c_{\downarrow\vect{k}'}c_{\uparrow(\vect{q}-\vect{k}')},
\label{eq:Hamiltonian}
\end{align}
where $\sigma\in\{\uparrow,\downarrow\}$ is the spin index, $\xi_{\sigma 
\vect{k}}$ is the single-particle dispersion for spin species $\sigma$ and 
momentum $\vect{k}$, $c^\dagger_{\sigma\vect{k}}$ ($c_{\sigma\vect{k}}$) is a 
creation (annihilation) operator for a fermionic particle, and $g>0$ is the 
strength of the attractive contact interaction.  In the absence of interactions
the 
ground state of the Hamiltonian in \eqnref{eq:Hamiltonian} is a filled Fermi 
sea,
\begin{align}
\ket{\text{FS}}=\prod_{\xi_{\uparrow \vect{k}_\uparrow}<\Efup}
c^\dagger_{\uparrow\vect{k}_\uparrow}
\prod_{\xi_{\downarrow \vect{k}_\downarrow}<\Efdn}
c^\dagger_{\downarrow\vect{k}_\downarrow}\ket{0},
\end{align}
with species-dependent Fermi energies $\Efsigma$ (corresponding to Fermi 
momenta $\kfsigma$) and $\ket{0}$ being the vacuum state.  Without loss of 
generality we fix the number of particles in the Fermi sea of the up-spin 
species to be greater than or equal to that of the down-spin species.  We 
follow the prescription of Cooper~\cite{Cooper56} and assume that the 
non-interacting ground state remains undisturbed for $\xi_{\sigma 
\vect{k}}<\Efsigma$, and focus only on a few-particle instability at 
$\xi_{\sigma \vect{k}}\gtrsim\Efsigma$.  We work in a general number $D\ge 2$ 
of dimensions.

\subsection{Fermi surface arcs}

The idealized conceptual situation where we expect a multi-particle instability 
to be present consists of two Fermi surfaces for the different species that are 
different sizes, but otherwise geometrically similar.  In \figref{fig:Circles} 
we show example Fermi surfaces for the up- and down-spin particles in $D=2$ 
dimensions.  Above the Fermi surfaces are the unoccupied momentum states that 
can host the multi-particle instability, which for typical phonon-mediated 
interactions extend over a species-independent Debye momentum $\kD$.  We assume 
that $\kD/\kfsigma\ll 1$, as for many conventional 
superconductors~\cite{Kok37,Reynolds51,Horowitz52}.

To construct the trial wavefunction for the multi-particle instability, we 
start by developing multi-particle basis states. To capture all possible 
correlations in the system, we require that the interaction term in the 
Hamiltonian can couple the different basis states.  As the interaction term 
conserves momentum, all basis states must have the same total momentum.  To 
construct these basis states, first consider the Cooper pair situation with 
only one up-spin and one down-spin particle in the instability.  We start with 
a basis state that has both particles on their respective Fermi surfaces on 
opposite sides of the Fermi seas (at the momenta labeled $\vect{q}_{\uparrow 
1}$ and $-\vect{q}_{\downarrow 1}$ in \figref{fig:IdealCirclesFFLO}).  In 
systems with anisotropic Fermi surfaces, like many of the candidate systems for 
FFLO~\cite{Aebi94,Shishido03,Lortz07}, the Cooper pair (and, later, the 
multi-particle instability) will be dominated by the lowest-curvature parts of 
the Fermi surface, and so in a general dispersion we place the initial basis 
state at the points on the Fermi surfaces with the lowest curvature.

If we move away from these starting momenta, tangentially to the Fermi surfaces 
by equal and opposite momenta for the different species to conserve momentum, 
we eventually reach the Debye momentum $\kD$ above the Fermi surfaces where 
there are no more momentum states accessible via the interaction term (reach 
the outer edge of the shaded regions in \figref{fig:IdealCirclesFFLO}).  The 
tighter curvature of the down-spin species means we will first run out of 
allowed momentum states for the down-spin species (at the point 
$-\vect{q}_{\downarrow 1}-\vect{k}_{\downarrow 1}$ in 
\figref{fig:IdealCirclesFFLO}).  The angular width of the allowed down-spin 
momentum states thus sets the angular width of the up-spin momentum states for 
Cooper pairs.

We refer to the allowed momentum states for the particles as forming `arcs' on 
the Fermi surfaces.  An idealized version of the available momentum states for 
the down-spin species is indicated in \figref{fig:IdealCirclesFFLO} by the arc 
above the down-spin Fermi surface bounded by blue lines, with angular width 
$\theta$.  The corresponding up-spin species arc is shown bounded by red lines.

Because it was the down-spin species that exhausted its available momentum 
states first in the Cooper pair situation in \figref{fig:IdealCirclesFFLO}, we 
wasted the opportunity for some up-spin species momentum states to become 
involved in the instability and so lower the energy of the system. We can make 
use of twice as many up-spin momentum states by duplicating the arc of up-spin 
momentum states that were available in the Cooper pair situation, offsetting 
the arcs so they do not intersect as required by Pauli exclusion, and placing a 
particle in each arc.  If we allow either up-spin particle to interact with the 
down-spin particle, we have increased the variational freedom in the system and 
would generically expect the binding energy to become larger.  Two such Fermi 
surface arcs for the up-spin species are shown in \figref{fig:IdealCircles}, 
bounded by red lines. 

We can generalize the above argument to include more than two up-spin and more 
than one down-spin particles: in general we may have $\Nup$ up-spin arcs and 
particles, and $\Ndn$ down-spin arcs and particles.  However, if we include too 
many particles, the gradients of the Fermi surfaces of the different species 
will differ radically at the extremal Fermi surface arcs, and it will not be 
possible to move around one species' arc without immediately pushing the other 
species out of their allowed momentum states.  We bound the maximum extent of 
the Fermi surface arcs by noting that when the tangents to the species' Fermi 
surfaces are parallel it is possible to move particles of both species 
simultaneously without either being forced from their allowed momentum states.  
For dispersions with inversion symmetry this is achieved when the total angular 
widths of the two species' arcs are equal, shown in \figref{fig:IdealCircles} 
with the total width of the arcs of both species taking the value $\theta$. 

The densities of states in momentum of the occupied arcs, $\nusigma$, describe 
the availability of momentum states throughout all arcs for each species, in 
our two-dimensional example being proportional to $\theta$.  The density of 
states in momentum per particle is then $\nusigma/\Nsigma$.  The species with 
the smaller value of this ratio limits the angular size available for each 
particle to move in, and we refer to this as the `critical' species, with $\Nc$ 
particles and density of states in momentum $\nuc$. 

To show that an instability with multiple particles in separate Fermi surface 
arcs is the energetically favorable solution for a broad class of 
spin-imbalanced systems, we follow the approach of Cooper~\cite{Cooper56} to 
construct a variational wavefunction for the multi-particle instability.  We 
demonstrate that in spin-imbalanced systems the multi-particle instability 
gives an improved binding energy over traditional Cooper pairs.

\subsection{Basis states}
\label{sec:bases}

To formalize the above description of the Fermi surface arcs, we label the 
angular center of each arc, on the Fermi surface, by a $q$-vector 
$\vect{q}_{\sigma i}$.  These $q$-vectors therefore satisfy $|\vect{q}_{\sigma 
i}|=\kfsigma$ and $\xi_{\sigma \vect{q}_{\sigma i}}=\Efsigma$.  For 
$\kD\ll\kfsigma$ and so small $\theta$ the $\vect{q}_{\sigma i}$ can be taken 
to be parallel, $|\vect{q}_{\sigma i}-\vect{q}_{\sigma j}|\ll|\vect{q}_{\sigma 
i}|$.  All the momenta within a particular arc are described by 
$\vect{q}_{\sigma i}+\vect{k}_{\sigma i}$, where the vectors $\vect{k}_{\sigma 
i}$ indicate the positions of the particles within the Fermi surface arcs, and 
for small $\kD\ll \kfsigma$ we have $|\vect{k}_{\sigma i}|\ll|\vect{q}_{\sigma 
i}|$.  This guarantees $\xi_{\uparrow(\vect{q}_{\uparrow i}+\vect{k}_{\uparrow 
i})}\gtrsim \Efup$ and $\xi_{\downarrow(-\vect{q}_{\downarrow 
j}-\vect{k}_{\downarrow j})}\gtrsim \Efdn$ so that the particle momenta lie 
near their corresponding Fermi surfaces. Examples of this labeling procedure 
are shown in \figref{fig:Circles}. 

The proposed multi-particle instability is an excitation of ($\Nup$,$\Ndn$) 
correlated particles on top of the undisturbed Fermi seas, with each particle 
existing in a unique arc.  This can be constructed out of basis states
\begin{align}
&\state{\Kup}{\Kdn}=\prod_{i}^{\Nup} 
c^\dagger_{\uparrow(\vect{k}_{\uparrow i}+\vect{q}_{\uparrow i})} 
\prod_{j}^{\Ndn} 
c^\dagger_{\downarrow(-\vect{k}_{\downarrow j}-\vect{q}_{\uparrow j})}
 \ket{\text{FS}},
\end{align}
where $\Ksigma=(\vect{k}_{\sigma 1},\vect{k}_{\sigma 2},\ldots,\vect{k}_{\sigma 
\Nsigma})$ is an $\Nsigma\times D$ matrix of particle momenta in $D$ spatial 
dimensions.

\subsection{Trial wavefunction}

The trial wavefunction for a system with a given set of $\vect{q}_{\sigma i}$ 
vectors is a sum over basis states with optimizable coefficients 
$\alpha(\Kup,\Kdn)$, 
\begin{align}
\ket{\psi}=\sideset{}{'}\sum_{\Kup,\Kdn}\alpha(\Kup,\Kdn)\state{\Kup}{\Kdn},
\label{eq:Wfn}
\end{align}
where the sum is over all $\Nsigma$ momentum components $\vect{k}_{\sigma i}$ 
of each matrix $\Ksigma$, with the prime on the sum indicating that we only sum 
over $\vect{k}_{\sigma i}$ such that $\sum_i^{\Nup}\vect{k}_{\uparrow 
i}=\sum_j^{\Ndn}\vect{k}_{\downarrow j}$, ensuring momentum conservation.  We 
take the $\alpha(\Kup,\Kdn)$ coefficients to be non-zero only if all of the 
momenta $\vect{k}_{\sigma i}$ lie within their respective arcs of the Fermi 
surfaces.  With $\Nup=\Ndn=1$ \eqnref{eq:Wfn} collapses to the trial 
wavefunction for a Cooper pair.

\subsection{Kinetic energy}

To find an analytic expression for the energy expectation value $E$, we first 
focus on the kinetic energy term and linearize the dispersions near the Fermi 
surfaces, $\xi_{\sigma \vect{p}}\approx (|\vect{p}|-\kfsigma) \xi_{\sigma 
\kfsigma}'$.  Here $\kfsigma$ is the momentum corresponding to the Fermi 
energy, which for small enough $\kD \ll \kfsigma$ can be considered constant 
over the Fermi surface arcs, and $\xi_{\sigma \kfsigma}'$ is the derivative of 
the single-particle energy at the Fermi surface.  For the dispersions involved, 
$\xi_{\uparrow (\vect{k}_{\uparrow i}+\vect{q}_{\uparrow i})}$ and 
$\xi_{\downarrow (-\vect{k}_{\downarrow i}-\vect{q}_{\downarrow i})}$, recall 
that $|\vect{q}_{\sigma i}|=\kfsigma$ and $|\vect{k}_{\sigma i}|\ll 
|\vect{q}_{\sigma i}|$, and so $\xi_{\uparrow (\vect{k}_{\uparrow 
i}+\vect{q}_{\uparrow i})} \approx k_{\uparrow i}\xi_{\uparrow \kfup}'$ and 
$\xi_{\downarrow (-\vect{k}_{\downarrow i}-\vect{q}_{\downarrow i})} \approx 
k_{\downarrow i}\xi_{\downarrow \kfdn}'$ where $k_{\sigma i}$ is the projection 
of $\vect{k}_{\sigma i}$ along $\vect{q}_{\sigma i}$ (or equivalently, the 
radial component of $\vect{k}_{\sigma i}$).

This linearity simplifies the full expression for the kinetic energy of our 
trial wavefunction.  With kinetic energy operator 
$\hat{T}=\sum_{\sigma,\vect{k}}\xi_{\sigma 
\vect{k}}c^\dagger_{\sigma,\vect{k}}c_{\sigma,\vect{k}}$, we find
\begin{align}
&\brastate{\Kup}{\Kdn}\hat{T}\ket{\psi}\neweqnline
&=\alpha(\Kup,\Kdn)\Bigg(\sum_i^{\Nup}\xi_{\uparrow (\vect{k}_{\uparrow i}+
\vect{q}_{\uparrow i})}+\sum_j^{\Ndn} \xi_{\downarrow (-\vect{k}_{\downarrow j}
-\vect{q}_{\downarrow j})}\Bigg)\neweqnline
&\approx\alpha(\Kup,\Kdn)\Bigg( \sum_i^{\Nup} k_{\uparrow i} 
\xi_{\uparrow \kfup}'+\sum_j^{\Ndn} k_{\downarrow j} \xi_{\downarrow \kfdn}'
\Bigg),
\end{align}
which may be simplified further by using the conservation of total momentum to 
define $\sum_i^{\Nup}k_{\uparrow i}=\sum_j^{\Ndn}k_{\downarrow j}= K$, giving
\begin{align}
\brastate{\Kup}{\Kdn}\hat{T}\ket{\psi}\approx 2\alpha(\Kup,\Kdn)K\xi',
\end{align}
with $\xi'=\frac{1}{2}(\xi_{\uparrow \kfup}'+\xi_{\downarrow \kfdn}')$.

\subsection{Potential energy}

To evaluate the total energy of the wavefunction $\ket{\psi}$ we also need to 
evaluate the effect of the potential energy operator 
\mbox{$\hat{V}=g\sum_{\vect{k},\vect{k}',\vect{q}}c^\dagger_{\uparrow(\vect{q}-
\vect{k})}c^\dagger_{\downarrow\vect{k}}c_{\downarrow\vect{k}'}c_{\uparrow(
\vect{q}-\vect{k}')}$}.  The interaction operator removes two particles, 
one of each 
spin species, from a basis state and then replaces them, having transferred 
momentum $\vect{m}=\vect{k}'-\vect{k}$ between them.  For a general basis state 
$\state{\Kup}{\Kdn}$ there are $\Nup\Ndn$ ways of choosing the pairs of 
particles that are involved.  We can formalize this procedure by defining
\begin{align}
\hat{V}\state{\Kup}{\Kdn}=g\!
\sum_{\vec{P}_\uparrow,\vec{P}_\downarrow,\vect{m}}\!
\state{\Kup-\vec{P}_\uparrow\otimes\vect{m}}
{\Kdn-\vec{P}_\downarrow\otimes\vect{m}},
\label{eq:potential}
\end{align}
and hence
\begin{align}
\brastate{\Kup}{\Kdn}\hat{V}\ket{\psi}=g \!\!
\sum_{\vec{P}_\uparrow,\vec{P}_\downarrow,\vect{m}}\!\!
\alpha\Big(\Kup-\vec{P}_\uparrow\otimes\vect{m},
\Kdn-\vec{P}_\downarrow\otimes\vect{m}\Big),
\end{align}
where the vectors $\vec{P}_\sigma$ form a set of standard basis vectors in 
particle-number space: each has one element that takes the value $1$, with the 
remaining $(\Nsigma-1)$ elements having value $0$.  These label the particles 
in the different arcs in \figref{fig:IdealCircles}.  The effect of an outer 
product of a $\vec{P}_\sigma$ vector with a scattering vector $\vect{m}$ is to 
construct the matrix $(\vecgrk{0},\ldots,\vect{m},\ldots,\vecgrk{0})$, where 
the column containing $\vect{m}$ is determined by the particular 
$\vec{P}_\sigma$ vector.  We sum over all possible pairs of up- and down-spin 
particles. 

\subsection{Multi-particle instability}
\label{sec:Instability}

We are now ready to combine the effect of the kinetic and potential energies by 
projecting the full Schr\"odinger equation $\hat{H}\ket{\psi}=E\ket{\psi}$ onto 
the state $\brastate{\Kup}{\Kdn}$ to calculate the energy expectation value 
$E$.  We find that
\begin{align}
&\left(2K\xi'-E\right)\alpha(\Kup,\Kdn)\neweqnline
&= g\sum_{\vec{P}_\uparrow,\vec{P}_\downarrow,\vect{m}}\alpha\Big(\Kup-
\vec{P}_\uparrow\otimes\vect{m},\Kdn-\vec{P}_\downarrow\otimes\vect{m}\Big),
\end{align}
which, following the approach of Cooper~\cite{Cooper56}, we divide by 
\mbox{$(2K\xi'-E)$} and sum over all $\Ksigma$ to obtain
\begin{align}
&\sideset{}{'}\sum_{\Kup,\Kdn}\alpha(\Kup,\Kdn)\neweqnline
&= g\sideset{}{'}\sum_{\Kup,\Kdn}\sum_{\vec{P}_\uparrow,\vec{P}_\downarrow,
\vect{m}}\frac{\alpha\Big(\Kup-\vec{P}_\uparrow\otimes\vect{m},\Kdn-
\vec{P}_\downarrow\otimes\vect{m}\Big)}{2K\xi'-E}.
\end{align}

Shifting the dummy momentum variables $\Ksigma$ on the right hand side by 
$\vec{P}_\sigma \otimes \vect{m}$ to remove the $\vec{P}_\sigma$ and $\vect{m}$ 
from the arguments of $\alpha(\Kup,\Kdn)$, we bring the implicit expression for 
the energy to the form
\begin{align}
&\sideset{}{'}\sum_{\Kup,\Kdn}\alpha(\Kup,\Kdn)\neweqnline
&= g\sum_{\vec{P}_\uparrow,\vec{P}_\downarrow}\sideset{}{'}\sum_{\Kup,\Kdn}
\alpha(\Kup,\Kdn)\sum_{\vect{m}}\frac{1}{2(K+m)\xi'-E},
\label{eq:ImpEnAlpha}
\end{align}
where $m$ is the radial projection of $\vect{m}$.  We can now separate the 
angular and radial parts of the sum over $\vect{m}$, and carry out the angular 
summation. The angular summation is limited by the critical species, giving a 
contribution of the density of available states $\nuc/\Nc$, meaning that the 
whole sum over $\vect{m}$ should be considered as over the critical species.

We can also make the substitution $m'=K+m$, which has the effect of restraining 
the $\Ksigma$ dependence of the right hand side of \eqnref{eq:ImpEnAlpha} 
entirely to the parameters $\alpha(\Kup,\Kdn)$ and the limits of the sums over 
$m'$. However, the momentum $m'$ accounts only for single momentum-transfer 
events, which following the prescription of Cooper theory have a maximum radial 
width in momentum of the Debye momentum $\kD$. 

The maximum kinetic energy $2m'\xi'$ of a basis state is obtained when each 
particle is at the upper end of its Fermi surface arc, giving a total kinetic 
energy \mbox{$2m'\xi'=(\Nup+\Ndn)\kD\xi'$}, and the minimum kinetic energy is 
obtained when each particle is at the bottom of its arc, for $2m'\xi'=0$.  The 
summation over $m'$ may be extended to cover this range, giving an implicit 
expression for the energy of
\begin{align}
&\sideset{}{'}\sum_{\Kup,\Kdn}\alpha(\Kup,\Kdn)\neweqnline
&=\frac{2g\nuc}{(\Nup+\Ndn)\Nc}\!\sum_{\vec{P}_\uparrow,\vec{P}_\downarrow}
\sideset{}{'}\sum_{\Kup,\Kdn}\!\alpha(\Kup,\Kdn)\!\!\!\!\!
\sum_{m'=0}^{\frac{(\Nup+\Ndn)\kD}{2}}\!\!\!\!\!\frac{1}{2m'\xi'-E}.
\label{eq:ImpEnAlphaP}
\end{align}
The only dependence on the $\Ksigma$ in the implicit expression for the energy 
is in the coefficients $\alpha(\Kup,\Kdn)$, so we can factorize out 
$\sideset{}{'}{\textstyle\sum}_{\Kup,\Kdn}\alpha(\Kup,\Kdn)$ from both sides of 
\eqnref{eq:ImpEnAlphaP}.  We have also removed all dependence on 
$\vec{P}_\sigma$ from the expression, and so can explicitly carry out those 
summations to give a factor of $\Nup\Ndn$.  This leaves us with
\begin{align}
1=g\frac{2\Nup\Ndn}{(\Nup+\Ndn)} \frac{\nuc}{\Nc} 
\sum_{m'=0}^{\frac{(\Nup+\Ndn)\kD}{2}}\frac{1}{2m'\xi'-E},
\label{eq:mintegral}
\end{align}
analogous to Eq.~(4) of Cooper's original paper~\cite{Cooper56}.

We have reduced the complexity of the multi-particle instability to a single 
summation with a multiplicative constant.  In the same manner as Cooper's 
original analysis we may now convert this summation to an integral and solve, 
finding the binding energy
\begin{align}
\Eb=\frac{(\Nup+\Ndn)\kD\xi'}{\exp\left(\frac{(\Nup+\Ndn)\xi'}
{g \Nup \Ndn}\frac{\Nc}{\nuc}\right)-1}.
\label{eq:BindingEnergy}
\end{align}
In the weakly interacting limit this binding energy simplifies to
\begin{align}
E_\mathrm{b}=(\Nup+\Ndn)\kD\xi'\exp\left(-\frac{(\Nup+\Ndn)\xi'}
{g \Nup \Ndn}\frac{\Nc}{\nuc}\right),
\label{eq:WeakBindingEnergy}
\end{align}
similar to the familiar form of the binding energy of a Cooper pair.

We wish to identify the number of particles $(\Nup,\Ndn)$ in the energetically 
optimal multi-particle instability.  The strongest dependence of the binding 
energy in \eqnref{eq:WeakBindingEnergy} on $\Nup$ and $\Ndn$ is in the 
exponential, with the binding energy being maximized when the argument of the 
exponential function is least negative.  The values of $\Nsigma$ which achieve 
this, and are therefore the optimal solutions for the system, can be deduced by 
symmetry to satisfy the relation
\begin{align}
\frac{\Nup}{\Ndn}=\frac{\nuup}{\nudn},
\label{eq:ratios}
\end{align}
i.e. the number of particles involved in the wavefunction per spin species is 
proportional to the density of states in momentum.  This means that all of the 
available momentum states are involved in the instability, and so contributing 
the maximum possible binding energy.  \eqnref{eq:ratios} suggests that in $D\ge 
2$ dimensions a multi-particle instability is energetically favorable over 
conventional pair instabilities in a spin-imbalanced system.  

In the next subsection we shall analyze our expression for the binding energy 
in the light of \eqnref{eq:ratios}, which gives a definite prediction for the 
energetically optimal instability in different systems.  We shall then return 
to the trial wavefunction given by \eqnref{eq:Wfn}, and look further at its 
properties and limits.

\subsection{Binding energy analysis}

To build our intuition for the expression for the binding energy of the 
multi-particle instability found in \eqnref{eq:BindingEnergy}, we now examine 
the binding energy as a function of the ratio of the number of particles 
$\Nup/\Ndn$. We render the binding energy dimensionless by normalizing by $g$, 
the interaction strength; $\kD$, the maximum interaction momentum; 
$\nuup\nudn$, in order to account for different system sizes; and $\Nc/\nuc$, 
the number of critical species particles per density of states in momentum.  
Normalizing by this final ratio looks forward to the eventual creation of a 
many-body strongly-correlated state from multi-particle instabilities, with the 
number of instabilities merged being limited by the availability of critical 
species particles.  We note, however, that at low interaction strengths the 
dominant term in the binding energy in \eqnref{eq:WeakBindingEnergy} is the 
exponential, so the normalization could be chosen to be by the total number of 
particles without affecting the results below.  This results in a measure of 
the binding energy per critical species particle of 
\begin{align}
X_\mathrm{b}=\frac{1}{g \kD \nu_\uparrow \nu_\downarrow}
\frac{E_\mathrm{b}}{N_\mathrm{c}/\nu_\mathrm{c}}.
\end{align}

To further justify this measure of the binding energy per critical species 
particle, we first examine the strongly interacting limit of 
\eqnref{eq:BindingEnergy}.  Here, in terms of the normalized ratio of number of 
particles per species,
\begin{align}
x=\frac{\Nup}{\Ndn}\frac{\nudn}{\nuup},
\end{align}
the binding energy per critical species particle takes the simple form
\begin{align}
X_\mathrm{b}=\left\{ \begin{matrix} x, & x<1, \\
1/x, & x>1.\end{matrix}\right.
\label{eq:idealhigh}
\end{align}
This expression is maximized at $x=1$, that is when $\Nup/\Ndn=\nuup/\nudn$, in 
agreement with the expression in \eqnref{eq:ratios} for the weakly-interacting 
limit.  

\begin{figure}
\includegraphics[width=\linewidth]{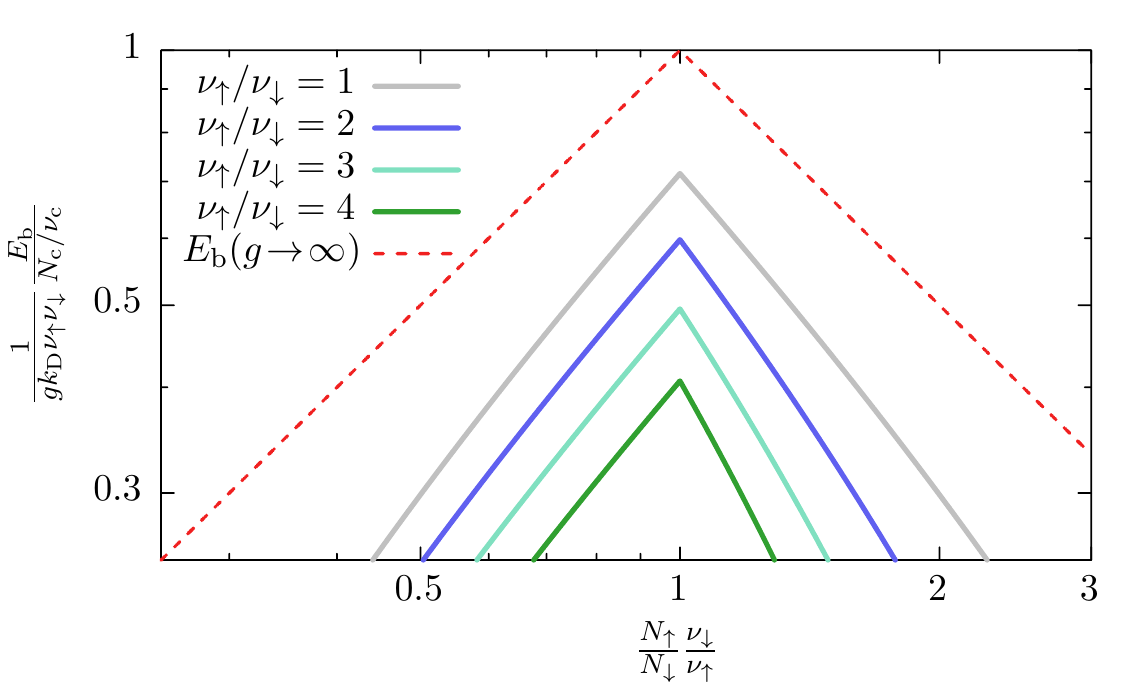}
\caption{(Color online) The binding energy per critical species particle as a 
function of the normalized ratio of number of particles per species at 
intermediate interaction strength $g=\Efup$.  Results are shown for a free 
dispersion in $D=2$ dimensions and for different imbalance ratios, with the 
infinite-interaction-strength limit indicated by a dashed line.}
\label{fig:energies}
\end{figure}

Away from the strongly- and weakly-interacting limits the optimal binding 
energy remains at $\Nup/\Ndn=\nuup/\nudn$.  In \figref{fig:energies} we show 
the binding energy per critical species particle $X_\mathrm{b}$ from 
\eqnref{eq:BindingEnergy} as a function of imbalance $x$ for ratios of 
densities of states in momentum $\nuup/\nudn\in\{1,2,3,4\}$ at an intermediate 
interaction strength $g=\Efup$.  We take as an example system a free dispersion 
with $\xi_{\sigma \kfsigma}'=\kfsigma$ in $D=2$ dimensions, so that 
$\nusigma\propto\kfsigma$, although similar results hold in other systems.  The 
balanced system $\nuup=\nudn$ is shown by the gray line, with the conventional 
Cooper state, having $\Nup=\Ndn$, being the energetically optimal instability.  
This line is symmetric about $\Nup/\Ndn=\nuup/\nudn$ on the log-log scale, 
which reflects the symmetry between spin species when $\nuup=\nudn$.  For the 
spin-imbalanced systems where $\nuup>\nudn$, the energetically optimal 
instability is still found at $\Nup/\Ndn=\nuup/\nudn$, as predicted by 
\eqnref{eq:ratios}.  To the right of this there are too many up-spin particles 
in the instability, and to the left there are too many down-spin particles in 
the instability; this leads to the $\nuup>\nudn$ lines not being symmetric 
about their maxima, as in imbalanced systems including the wrong number of 
up-spin particles is not equivalent to including the wrong number of down-spin 
particles.

Having examined the result of \eqnref{eq:ratios} that the optimum ratio of 
number of particles is given by the ratio of densities of states in momentum, 
we now discuss the difference between instabilities with different numbers of 
particles, but the same ratio $\Nup/\Ndn$.

\subsection{Instabilities with same ratio $\Nup/\Ndn$}
\label{sec:ratios}

The prediction given in \eqnref{eq:ratios} that the energetically optimal 
numbers of particles involved in the instability are related by 
$\Nup/\Ndn=\nuup/\nudn$ only sets the ratio between $\Nup$ and $\Ndn$, but does 
not predict the absolute numbers of particles.  To probe the effect of changing 
the absolute numbers of particles, we need to examine in more detail the effect 
of Pauli blocking.

The effect of Pauli blocking has been carefully 
analyzed~\cite{Pogosov10,Pogosov11,Fan10} for the product of two 
$(\Nup,\Ndn)=(1,1)$ instabilities, and found to give only a small correction to 
the binding energy of two separate pairs (the correction going as the inverse 
of the number of available momentum states).  This agreement with our result 
for a $(\Nup,\Ndn)=(2,2)$ instability, up to small Pauli blocking corrections 
that vanish in the thermodynamic limit, supports our finding that the binding 
energy per critical species particle is independent of the total number of 
particles involved in the instability.  We shall present further numerical 
evidence that captures the Pauli blocking corrections in 
Subsection~\ref{sec:ExDiagRatio}.

However, the effect of Pauli blocking will become more acute in a many-body 
state constructed from multi-particle instabilities.  This suggests that in the 
limit of a large number of multi-particle instabilities in a system, 
instabilities with fewer total particles will be energetically favorable over 
instabilities with more particles but the same value of $\Nup/\Ndn$ in a given 
system.

Having investigated the structure and binding energy of the proposed 
multi-particle instability, we now turn to some of its limits.  We examine the 
conventional Cooper system, with balanced Fermi seas, and identify the 
predictions made for one-dimensional systems, recovering in both cases 
agreement with well-known results from the literature. We also briefly examine 
the strongly-interacting limit of the proposed multi-particle instability.

\subsection{Cooper limit}

The system studied originally by Cooper~\cite{Cooper56} is a balanced Fermi 
gas, and so has $\kfup=\kfdn=\kf$, $\nuup=\nudn=\nu$, which we predict should 
have the optimal ratio $\Nup/\Ndn=1$ in agreement with Cooper's findings.  
Moreover, with $\Nup=\Ndn=1$ our trial wavefunction \eqnref{eq:Wfn} reproduces 
the conventional Cooper trial wavefunction~\cite{Cooper56}.  Therefore, with a 
free dispersion $\xi_{\sigma k}=k^2/2-\kf^2/2$ the weakly-interacting binding 
energy given by \eqnref{eq:WeakBindingEnergy} reduces to the familiar Cooper 
expression~\cite{Cooper56}
\begin{align}
\Eb=2\kD\xi' \exp\left(-\frac{2\kf}{g}\frac{1}{\nu}\right)
=2\omegaD \exp\left(-\frac{2}{g\Omega}\right),
\end{align}
where the Debye energy $\omegaD=\kD\xi'$ and the density of states in energy 
$\Omega=\nu/\kf$.

\subsection{One-dimensional limit}

Although the discussion in previous subsections has focused on $D\ge 2$ 
dimensions, our main prediction of $\Nup/\Ndn=\nuup/\nudn$ also holds in $D=1$ 
dimension.  Here the density of states in momentum is independent of the Fermi 
momentum, and so $\nuup/\nudn=1$ for both balanced and imbalanced systems.  
This suggests that a Cooper pair instability with $\Nup=\Ndn=1$ should be 
energetically optimal for both balanced and imbalanced systems in $D=1$ 
dimension.  This is in agreement with both analytical 
predictions~\cite{Yang01,Hu07,Parish07} and numerically exact 
calculations~\cite{Feiguin07,Batrouni08} that show an FFLO phase constructed 
from Cooper pairs is the ground state throughout a large part of the phase 
diagram of one-dimensional Fermi gases.

\subsection{Strongly-interacting limit}

In the limit of strong attractive interactions $g\gg \Efsigma$ we expect the 
system to promote particles to the energy of the 
up-spin Fermi surface to reconstruct full rotational symmetry, similar to a 
breached superconductor~\cite{Gubankova03,Liu03,Parish07a}.  This turns the 
system effectively into one with balanced reconstructed Fermi surfaces, and so 
supporting conventional Cooper pair instabilities.  In the strongly-interacting 
limit of a many-body theory built from Cooper pairs, the pair coherence length 
becomes small on the scale of the separation between pairs, and so the pairs 
can be considered tightly-bound dimers~\cite{Massignan14,Nozieres85}.

We have shown that the proposed multi-particle instability reduces to the 
well-studied Cooper problem in the balanced limit, and collapses to a pair 
instability in one dimension, which both link with previous results, and 
also reproduces a known result in the strongly-interacting limit.
This gives us confidence that the multi-particle 
construction is also valid away from these limits. Having shown the strength of 
the formalism in reproducing these known limits, we now provide numerical 
evidence for the multi-particle instability being energetically optimal in a 
range of spin-imbalanced systems.

\section{Exact diagonalization}
\label{sec:ExDiag}   

\begin{figure}[t]
\includegraphics[width=\linewidth]{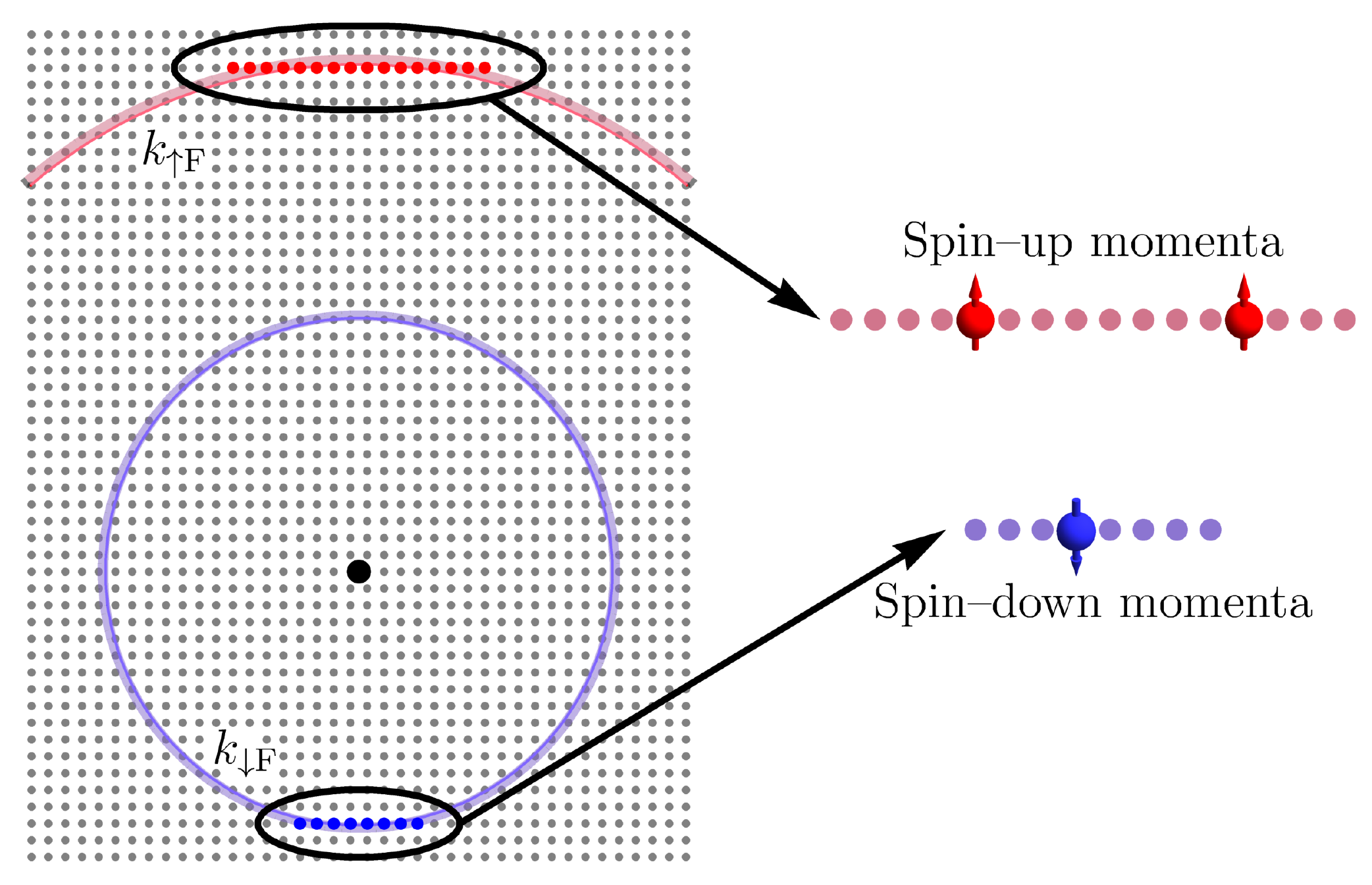}
\caption{(Color online) Example discretized momentum states (gray points) for 
use in exact diagonalization calculations, showing up- and down-spin Fermi 
surfaces (red and blue curves) with a ratio of $\nuup/\nudn=2$.  The origin is 
marked by the large black point. The subsets of momentum states used in 
calculations are colored and circled, in this case showing a 
$(\Sup,\Sdn)=(16,8)$ system.  These states are shown larger, for clarity, on 
the right-hand side, with sample particle occupations.}
\label{fig:ExDiagPoints}
\end{figure}

\subsection{Method}

In order to provide further insights into our conclusion that the optimal ratio 
of number of particles in an instability is given by $\Nup/\Ndn=\nuup/\nudn$ we 
turn to a numerical evaluation of the wavefunction $\ket{\psi}$ and energy 
expectation value $\bra{\psi}\hat{H}\ket{\psi}/\braket{\psi}{\psi}$.  To gain 
computational traction, we examine a reduced Hilbert space, taking only 
finitely many momentum states from the Fermi surfaces.  We indicate this 
reduction in Hilbert space size in \figref{fig:ExDiagPoints}, where instead of 
considering all momentum states (gray points) or even all momentum states on 
the up- and down-spin Fermi surfaces (red and blue curves), we use just linear 
subsets from opposite sides of the Fermi surfaces.  This allows us to focus on 
the angular extent of the Fermi arcs, the driving force behind the emergence of 
the multi-particle instability.  We work in the strongly interacting limit, to 
minimize the effect of neglecting the radial component of the sum over 
momentum.  We use systems with $\Sup$ momentum states for up-spin particles, 
and $\Sdn$ momentum states for down-spin particles: the ratio $\Sup/\Sdn$ then 
mimics the ratio of densities of states in momentum $\nuup/\nudn$. 
\figref{fig:ExDiagPoints} shows an example system with $(\Sup,\Sdn)=(16,8)$.

To numerically identify the ground state of the $(\Nup,\Ndn)$ system of 
particles in a system with $(\Sup,\Sdn)$ momentum states, we explicitly 
construct the $\binom{\Ssigma}{\Nsigma}$ combinations of particle momenta for 
each species, for a total of $\binom{\Sup}{\Nup}\times\binom{\Sdn}{\Ndn}$ basis 
states.  Note that we do not explicitly include the additional constraint of 
the separation into Fermi surface arcs used in the wavefunction 
\eqnref{eq:Wfn}.  We then directly evaluate and diagonalize the matrix of 
interactions between these states, with the optimal instability being that with 
the most negative eigenvalue.

\subsection{Binding energy}

\begin{figure}[t]
\includegraphics[width=\linewidth]{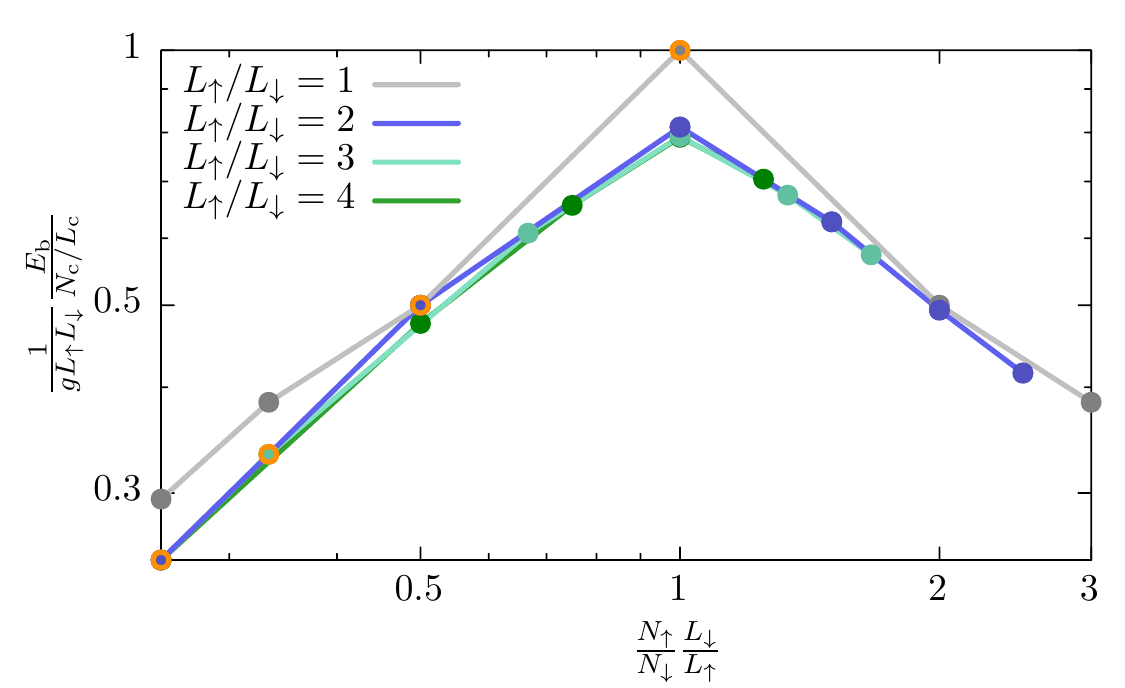}
\caption{(Color online)  The normalized binding energy per critical species 
particle obtained using exact diagonalization (lines), including the Cooper 
pair values (orange points).}
\label{fig:EDenergy}
\end{figure}

We investigate the dependence of the optimal binding energy on the ratio of 
number of particles $\Nup/\Ndn$ in the instability in \figref{fig:EDenergy}, 
where we plot the normalized binding energy per critical species particle 
against the ratio of number of particles per species, normalized by the inverse 
ratio of number of momentum states.  This rescaling of $\Nup/\Ndn$ ensures that 
our predicted optimal binding energies are located at $\Nup\Sdn/\Ndn \Sup=1$, 
as in \figref{fig:energies}. We examine systems with different ratios of 
numbers of momentum states $\Sup/\Sdn\in\{1,2,3,4\}$, with the lines in 
\figref{fig:EDenergy} coming from systems containing $(\Sup,\Sdn)=(16,16)$, 
$(\Sup,\Sdn)=(16,8)$, $(\Sup,\Sdn)=(18,6)$, and $(\Sup,\Sdn)=(16,4)$ momentum 
states respectively.

We observe that, as predicted by \eqnref{eq:ratios}, the optimal binding energy 
per critical species particle for each ratio of number of momentum states is 
obtained with a ratio of number of particles of $\Nup/\Ndn=\Sup/\Sdn$.  This is 
the principal result of our exact diagonalization investigation: our numerical 
study reproduces the result of our approximate analytical method.

To highlight that Cooper pairs are suboptimal in spin-imbalanced systems, we 
indicate the Cooper pair instability for each system in \figref{fig:EDenergy} 
with orange circles, from left to right for the $\Sup/\Sdn=4$, $\Sup/\Sdn=3$, 
$\Sup/\Sdn=2$, and $\Sup/\Sdn=1$ systems.  We note that for $\Sup/\Sdn>1$ these 
Cooper pair states have lower binding energy per critical species particle than 
the proposed multi-particle instability, whilst for $\Sup=\Sdn$ the optimal 
multi-particle instability is simply a Cooper pair, as predicted by 
Cooper~\cite{Cooper56}.

\begin{figure}[t]
\includegraphics[width=\linewidth]{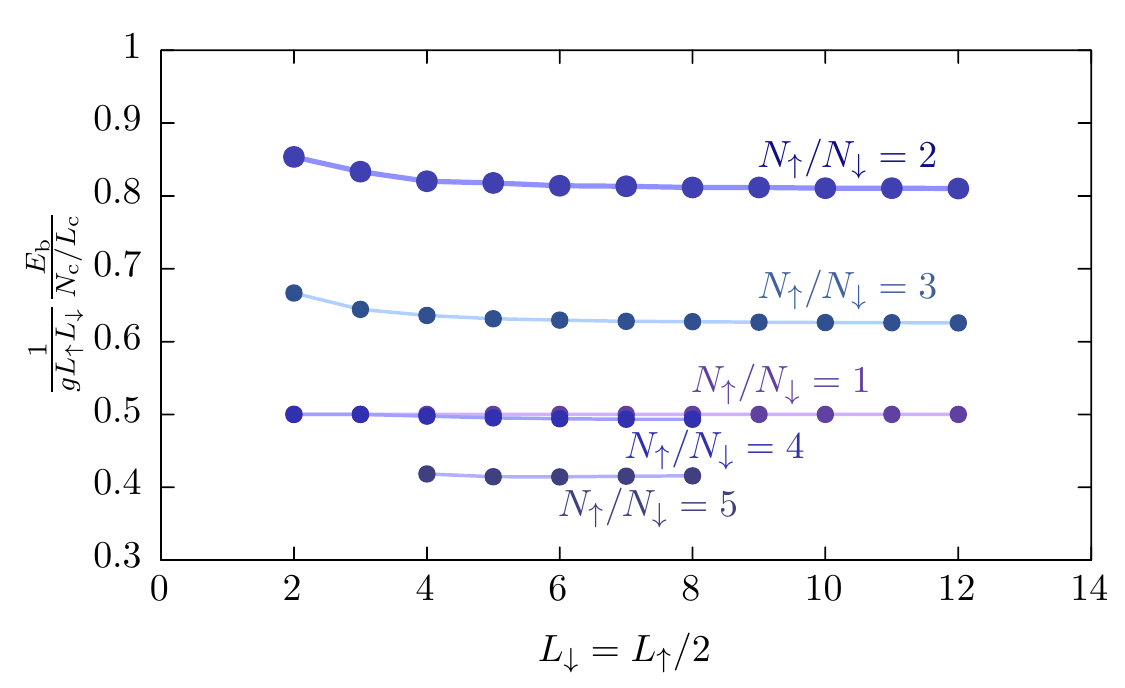}
\caption{(Color online) System size dependence of the binding energy per 
critical species particle, for ratio of number of momentum states 
$\Sup/\Sdn=2$.  The different lines correspond to different ratios of numbers 
of particles in the instabilities.}
\label{fig:FinSize}
\end{figure}

In \figref{fig:FinSize} we confirm the convergence of our exact diagonalization 
results with respect to system size for an example ratio $\Sup/\Sdn=2$. The 
different blue lines in \figref{fig:FinSize} correspond to exact 
diagonalization calculations of the binding energy using different numbers of 
up-spin particles, $\Nup\in\{ 1,2,3,4,5 \}$, and fixed $\Ndn=1$.  We observe a 
rapid convergence to the infinite size limit, with the $(\Sup,\Sdn)=(16,8)$ 
system shown in \figref{fig:ExDiagPoints} giving results within $0.4$\% of the 
infinite size limit for the $\Nup/\Ndn\in\{1,2,3\}$ ratios of numbers of 
particles. A slice through \figref{fig:FinSize} at $\Sdn=8$ gives the line for 
$\Sup/\Sdn=2$ in \figref{fig:EDenergy}.

\subsection{Fermi surface arcs}

\begin{figure}[t]
\includegraphics[width=\linewidth]{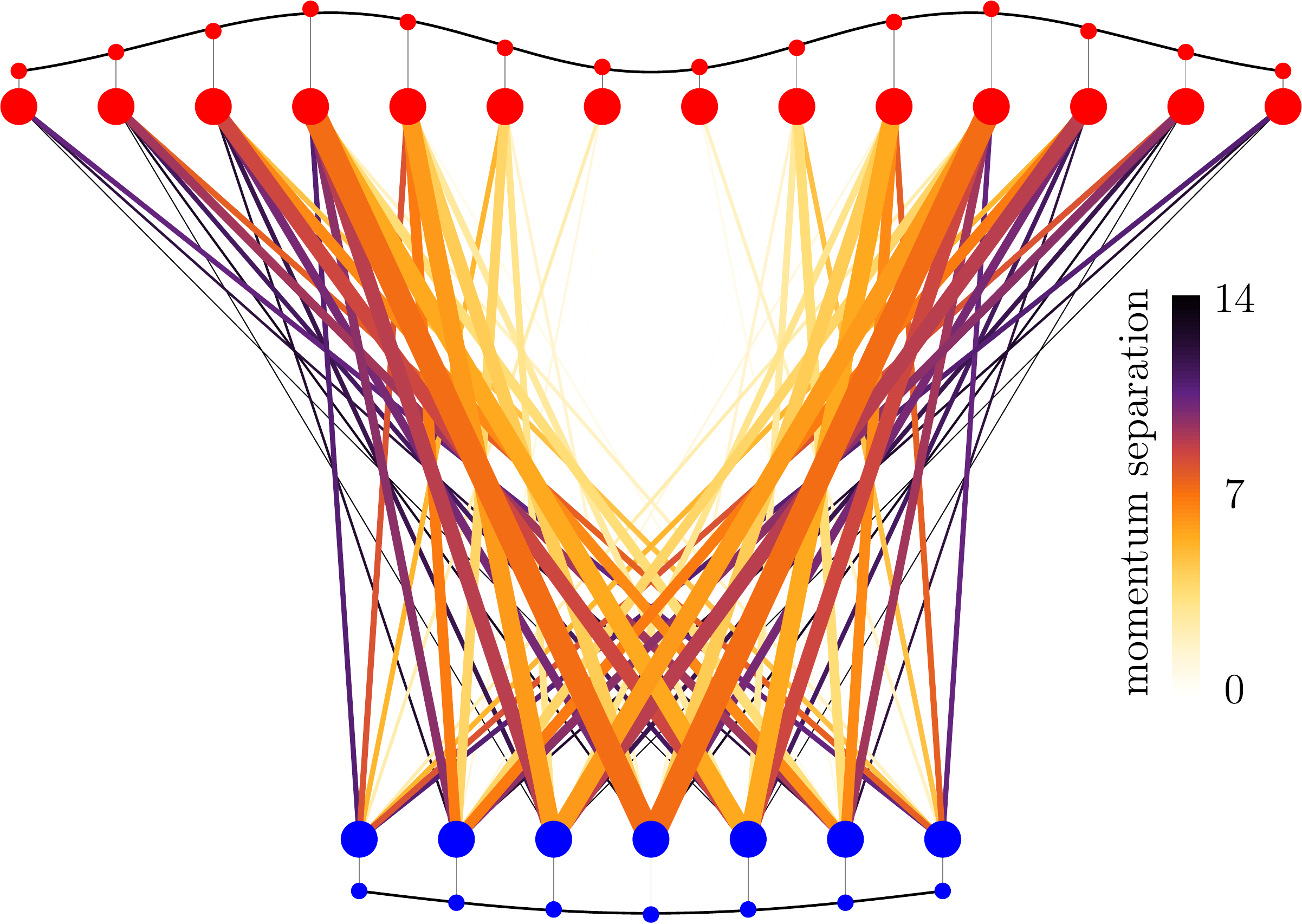}
\caption{(Color online) The weighting of basis states for the 
$(\Nup,\Ndn)=(2,1)$ instability of the $(\Sup,\Sdn)=(14,7)$ system.  Colored 
lines between the momentum states (represented by large red (up-spin) and blue 
(down-spin) points) indicate the basis states: each down-spin momentum state is 
part of a basis state with the two up-spin momentum states joined to it by 
lines of the same color and thickness.  Thicker lines indicate higher weighting 
(larger $\alpha(\Kup,\Kdn)$) of the basis states; thinner lines indicate lower 
weighting.  Color indicates the separation of the up-spin momentum states in 
each basis state, with yellow indicating small separation and purple large 
separation, with the color key indicating the separation in number of momentum 
states.  Only the $35$ highest weighted basis states are shown for clarity.  
Above the up-spin momentum states and below the down-spin momentum states are 
the integrated weights of the basis states at each momentum state, indicating 
the separation of the momentum states into Fermi surface arcs.}
\label{fig:EDoccupation}
\end{figure}

It is also illuminating to examine the wavefunctions of the energetically 
optimal instabilities.  In \figref{fig:EDoccupation} we show the basis states 
involved in the energetically optimal $(\Nup,\Ndn)=(2,1)$ instability of the 
$(\Sup,\Sdn)=(14,7)$ system.  Each down-spin momentum state is part of a basis 
state with the two up-spin momentum states joined to it by lines of the same 
thickness and color.  Thicker lines indicate higher weighting (larger 
$\alpha(\Kup,\Kdn)$) of the basis states, and colours represent the separation 
in momentum between the up-spin species particles in the instability. The 
wavefunction comprises basis states that have spontaneously organized arcs of 
the up-spin Fermi surface: each plotted basis state has one up-spin particle in 
the left-hand half of the up-spin Fermi surface, and one particle in the 
right-hand half.  This is in agreement with the use of arcs in the analytical 
wavefunction given by \eqnref{eq:Wfn}. In addition, the highest-weighted basis 
states are those at the angular center of the arcs, which are the momenta 
labeled $\vect{q}_{\sigma i}$ in \secref{sec:Theory}.

The separation of the wavefunction into Fermi surface arcs is also indicated by 
the integrated weights of the basis states at each momentum state, which are 
shown by the small points above the up-spin momentum states and below the 
down-spin momentum states in \figref{fig:EDoccupation}.  The integrated weights 
for the up-spin particles show a bimodal distribution, indicating a separation 
into arcs.  The black lines are symmetric fits to the data points, showing the 
arcs to contain identical distributions of integrated weights.  As expected for 
an $\Ndn=1$ system, the down-spin particle inhabits a single Fermi surface arc.

\subsection{Instabilities with same ratio $\Nup/\Ndn$}
\label{sec:ExDiagRatio}

Exact diagonalization may also be used to confirm the conclusion of 
Subsection~\ref{sec:ratios} that instabilities with fewer total particles are 
marginally energetically favorable over instabilities with the same ratio 
$\Nup/\Ndn$, but more particles.  By examining the binding energy per particle 
of the simple $(\Nup,\Ndn)=(1,1)$ and $(\Nup,\Ndn)=(2,2)$ instabilities in 
balanced systems with $\Sup=\Sdn=L$, we observe that the $(\Nup,\Ndn)=(1,1)$ 
instability does indeed have higher binding energy per particle at all finite 
interaction strengths.  As predicted analytically~\cite{Pogosov10}, the 
difference scales as $L^{-1}$ in the weakly-interacting limit, confirming the 
conclusion that instabilities with fewer total particles are energetically 
favorable in finite systems.

Our exact diagonalization results on this simplified system have supported the 
main claims and conclusions of the analytical arguments in \secref{sec:Theory}. 
 The energetically optimal instability in a range of different spin-imbalanced 
systems has been shown to satisfy the relationship $\Nup/\Ndn=\Sup/\Sdn$ 
predicted in \eqnref{eq:ratios}.  The separation of the Fermi surface into arcs 
in the analytical wavefunction has also been justified by the emergence of such 
arcs in the numerical calculations, and we have provided evidence for which 
instabilities with the same ratio $\Nup/\Ndn$ are most energetically favorable.

\section{Discussion}

We have shown that spin-imbalanced Fermi gases with attractive interactions 
support a multi-particle instability.  The most energetically favorable 
instability contains up- and down-spin particles in the ratio 
$\Nup/\Ndn=\nuup/\nudn$, set by the ratio of the densities of states in 
momentum at the Fermi surfaces. 

The proposed trial wavefunction for the multi-particle instability interpolates 
between the well-known Cooper wavefunction~\cite{Cooper56} in the limit of 
balanced Fermi surfaces and theoretical predictions~\cite{Yang01,Hu07,Parish07} 
of the FFLO phase in one dimension.  This lends support to the contention that 
our trial wavefunction is also valid away from these limits.

We note that the physics presented here can be explored in few-body systems.  
Cold atoms in an harmonic trap~\cite{Serwane11,Zurn12} are an ideal system to 
explore few-particle physics, as the exact energy and expectation values such 
as the wavefunction symmetry may be directly 
measured~\cite{Bugnion13,Bugnion13a}.  Cold atom experiments may therefore be 
able to observe the scaling of the binding energy and spatial structure of the 
trial wavefunction proposed here, of which hints may previously have been seen 
numerically~\cite{Bugnion13a}.

In real experiments the interaction between fermions will never be exactly the
contact interaction from \eqnref{eq:Hamiltonian}.  In cold atom systems the 
interaction may be expanded as $g(1+8a R_\mathrm{eff}\kf^2)$, where $a<0$ is
the scattering length and $R_\mathrm{eff}$ is the effective 
range~\cite{Keyserlingk13}.  Positive $R_\mathrm{eff}$ reduces the effective
interaction strength, making the multi-particle instability less energetically
favorable, whilst negative $R_\mathrm{eff}$ makes it more energetically
favorable; however, $|R_\mathrm{eff}|$ is typically small on the scale
of $1/\kf$, and so the effect of the finite range interaction is also small. 
The screened Coulomb interaction, $g/(1+2b^2\kf^2)$ where $b>0$ is the 
Thomas-Fermi screening length,
relevant for example to electron-hole systems~\cite{Lozovik75,Perali13}, 
has a similar effect, with the screening length taking on the same role as the 
effective range for cold-atom interactions, and so weakening the multi-particle 
instability relative to the purely-contact case.  This weakening is also found
in standard Cooper pairs, however, and so is unlikely to qualitatively change
the conclusions in the manuscript.  The next order
term in the effective range expansion would go as $R_\mathrm{eff}\kD$: this
term will discriminate between multi-particle instabilities
and Cooper pairs, being a function of how many fermions near the Fermi surfaces
are involved in the instability, but is not expected to have a large effect, as
 in our formalism both $\kD$ and $R_\mathrm{eff}$ are small.

In the same way that Cooper pairs form the conceptual basis of the 
Bardeen--Cooper--Schrieffer theory of superconductivity, it is expected that a 
many-body state may be constructed using the multi-particle instabilities 
presented here with even values of $\Nup+\Ndn$.  By analogy to the relationship 
between the traditional Cooper result and the BCS order parameter, we expect 
that the order parameter of the future many-body superconducting theory should 
have a form that is reminiscent of \eqnref{eq:WeakBindingEnergy}.  The 
many-body theory should not be limited to including a single type of 
multi-particle instability, and similarly to predictions made for the FFLO 
phase~\cite{Bowers02} may be constructed from multiple superposed 
multi-particle instabilities, forming a crystalline structure.

A natural tool to use to search for this exotic superconducting state is a 
spin-imbalanced ultracold fermionic gas~\cite{Zwierlein06,Partridge06}.  This
system allows fine control over the populations and interactions of the 
fermions, allowing experiments to focus on the potential of new physics.  
Previous spin-imbalanced ultracold fermionic gas experiments have used
inhomogeneous optical trapping potentials, in which the region of space where
multi-particle instability-based superconductivity is likely to be observable 
is very small.  However, recent experimental developments have
allowed the creation of homogeneous ultracold fermionic 
gases~\cite{Mukherjee17}, where the delicate novel superconducting state is 
likely to exist over larger regions of space, and so be easier to observe and
characterize.

Such a strongly correlated state would present novel superconducting 
properties, including unusual Andreev reflection~\cite{Andreev64}, Josephson 
tunneling~\cite{Josephson62}, and SQUID~\cite{Jaklevic64} or other 
superconducting loop~\cite{Kirsanskas15} properties, due to the underlying 
multi-particle structure.  With the underlying instabilities involving 
$\Nup+\Ndn$ fermions, magnetic flux is likely to be quantized in units of 
$h/(\Nup+\Ndn)e$, rather than $h/2e$ for BCS superconductivity based on 
Cooper pairs.  The superconducting order parameter would also
exhibit unusual behavior, being necessarily complex due to the presence of 
non-antipodal $q$-vectors, and oscillating with wavevectors 
$\vect{q}_{\uparrow i}+\vect{q}_{\downarrow j}$, with interference due to 
similar $q$-vectors giving rise to beats in the order parameter amplitude.  The
existence of a superconducting state constructed from multi-particle 
instabilities may also explain 
the lack of definitive observations of the conjectured FFLO state.

Data used for this paper are available online~\cite{data}.

\begin{acknowledgments}
  The authors thank Adam Nahum, Johannes Hofmann, Johannes Knolle, Jens Paaske, 
Robin Reuvers, and Darryl Foo for useful
  discussions, and acknowledge the financial support of the EPSRC and the Royal 
Society.
\end{acknowledgments}

\end{document}